\documentclass[10pt,twocolumn,nofootinbib,showpacs,aps,prd]{revtex4-1}
\usepackage{amsfonts,amsmath,amssymb,bm,mathrsfs}
\usepackage{mathtools,mathptmx,array,latexsym}
\usepackage[colorlinks=true,linkcolor=blue,citecolor=blue,urlcolor=blue]{hyperref}
\usepackage{url}
\usepackage{color} 

\def\bea{\begin{eqnarray}}
\def\eea{\end{eqnarray}}
\def\be{\begin{equation}}
\def\ee{\end{equation}}
\def\beal{\begin{aligned}}
\def\eeal{\end{aligned}}
\def\nn{\nonumber}
\def\p{\partial}

\begin{document}

\title{Notes on thermodynamics of super-entropic AdS black holes}

\author{Di Wu$^{1,2}$}
\author{Puxun Wu$^{1}$}
\email{Corresponding author, \\ pxwu@hunnu.edu.cn}
\author{Hongwei Yu$^{1}$}
\author{Shuang-Qing Wu$^{2}$}

\affiliation{$^{1}$Department of Physics and Synergetic Innovation Center for Quantum
Effects and Applications, Hunan Normal University, Changsha, Hunan 410081,
People's Republic of China \\
$^{2}$College of Physics and Space Science, China West Normal University, Nanchong,
Sichuan 637002, People's Republic of China}

\date{\today}

\begin{abstract}
The super-entropic black hole, which possesses a noncompact horizon topology and violates
the reverse isoperimetric inequality, has been found to satisfy both the thermodynamic
first law and the Bekenstein-Smarr mass formula. In this paper, we first derive a new
Christodoulou-Ruffini-like squared-mass formula for the four-dimensional Kerr-Newman-AdS
super-entropic black hole, and then establish a set of very simple relations between
thermodynamic quantities of the super-entropic Kerr-Newman-AdS$_4$ black hole and
its usual counterparts. Using these relations, the thermodynamic quantities of the
Kerr-Newman-AdS$_4$ super-entropic black hole can be obtained from those of the usual
pro-type by taking the ultra-spinning limit properly. Then these relations are extended
to the singly-rotating Kerr-AdS black holes in arbitrary dimensions and the double-rotating
charged black hole in the five-dimensional minimal gauged supergravity. It can be inferred
that the thermodynamic quantities of all super-entropic black holes obey similar limiting
relations to those of their corresponding conventional rotating AdS black holes, and
thus can be obtained by taking the ultra-spinning limit appropriately.
\end{abstract}


\maketitle

\section{Introduction}

Black hole is the most fundamental and important object predicted by Einstein's general
relativity. Constructing exact black hole solutions to the Einstein equation and studying
their properties can deepen our understanding of the nature of gravity and the basic
property of spacetime. Recently, a new class of the so-called super-entropic black
hole \cite{PRL115-031101,PRD89-084007}, which provides the first example that violates
the ``reverse isoperimetric inequality'' \cite{PRD84-024037,PRD87-104017}, has received
considerable interest and enthusiasm. This kind of black hole solution is asymptotical
(locally) anti-de Sitter (AdS) and has a finite horizon area, with its horizon topology
being noncompact since its topological sphere has two punctures, one at the north and
the other at the south.

Remarkably, it has been shown \cite{PRL115-031101} that the super-entropic black hole solution
can be alternatively obtained by simply taking the ultra-spinning limit of the usual rotating
AdS one. Taking the four-dimensional Kerr-Newman-AdS black hole or an arbitrary dimensional
Kerr-AdS black hole as an example, the concrete procedure to obtain the super-entropic
black hole is as follows: First rewrite the rotating AdS black hole in a rotating frame
at infinity, then boost the rotating angular velocity (only one rotating angular velocity
if there exist several rotating axes) to the speed of light, and finally compactify the
corresponding azimuthal direction. However, it was claimed \cite{PRL115-031101,JHEP0615096}
that the thermodynamic quantities of the super-entropic black hole can not be obtained from
those of its corresponding usual rotating AdS black hole by simply taking the ultra-spinning
limit. This is simply because some thermodynamic quantities will be divergent or zero when the
ultra-spinning limit is taken directly. So far, the thermodynamic quantities of super-entropic
black hole have been obtained usually by using the standard calculation method \cite{PRL115-031101,
JHEP0615096,NPB903-400,PRD95-046002,JHEP0118042,1702.03448}. As such, a question arises as to
whether there are certain relations between the thermodynamic quantities of the usual rotating
AdS black hole and their corresponding super-entropic ones, and whether the thermodynamic
quantities of the super-entropic black hole can be appropriately derived from those of its usual
rotating AdS black hole by taking the ultra-spinning limit. In this paper, we will propose
a new, simple way to relate them when the ultra-spinning limit is properly taken. Once
this is done, one can greatly simplify the calculations and step toward discussing the
super-entropic black hole thermodynamics. Very recently, Appels \textit{et al}. presented
a different method that the super-entropic black hole can also be obtained by running a
conical deficit through the usual rotating AdS black hole \cite{1911.12817}.

It is well known that thermodynamics of a black hole include three famous mass formulas, i.e.,
the first law of thermodynamics \cite{PRD7-2333,PRD13-191}, the Bekenstein-Smarr mass formula
\cite{PRL30-71} and the Christodoulou-Ruffini squared-mass formula \cite{PRL25-1596,PRD4-3552}.
Quite recently, these formulas were perfectly extended \cite{PRD100-101501} to the four-dimensional
NUT-charged spacetimes. It has been found \cite{PRL115-031101} that the first law of thermodynamics
and the Bekenstein-Smarr mass formula can be established for the super-entropic black hole if
an extra thermodynamic conjugate pair is introduced. However, a similar Christodoulou-Ruffini
squared-mass formula is still absent until now. Actually, the existence of such a squared-mass
formula is very important, since it can be used to check whether a new conjugate pair introduced in
the first law and the Bekenstein-Smarr mass formula is correct or not. In addition, recent
studies have shown that the squared-mass formula satisfied by a black hole can be used to
study black hole chemistry \cite{CQG26-195011,PRD92-124069,CQG34-063001}, thermodynamic Ruppeiner geometry
\cite{RMP67-605,PRD75-024037,PRD78-024016} and phase transitions \cite{PRL123-071103,1909.03887,
1909.11911,1910.07874,1910.03378,1910.04528}. Thus, it is desirable to get a similar
Christodoulou-Ruffini-like squared-mass formula for the super-entropic black hole too.

In this paper, we first establish a novel Christodoulou-Ruffini-like squared-mass formula for the
four-dimensional Kerr-Newman-AdS super-entropic black hole. Differentiating this formula gives the
thermodynamic quantities of this super-entropic black hole, which satisfy both the first law and
the Bekenstein-Smarr mass formula. Then, we construct a set of simple relations between thermodynamic
quantities of the super-entropic black hole and those of its usual counterpart. Basing upon these
relations, the thermodynamic quantities of the super-entropic black hole can be obtained from those
of its usual rotating AdS black hole by taking the ultra-spinning limit appropriately. The remaining
part of this paper is organized as follows. In Section 2, we first present a brief review of the
thermodynamical properties of the Kerr-Newman-AdS$_4$ super-entropic black hole, and then construct
a new Christodoulou-Ruffini-like squared-mass formula for it,   from which both the differential
and integral mass formulas can be derived via a simple mathematical manipulation without taking into
account the chirality condition ($J = Ml$). Then, we discuss the impact of the chirality condition on the
mass formulas and the super-entropic black hole thermodynamics.  After that,  we establish a set of simple
relations between thermodynamic quantities of the super-entropic Kerr-Newman-AdS$_4$ black hole and
its usual counterparts, by which we can straightforwardly obtain the thermodynamic quantities of the super-entropic
black hole when the ultra-spinning limit is taken properly. In Section 3, we turn to discuss the
case of the singly-rotating Kerr-AdS black holes in arbitrary dimensions. Then in Section 4, we
extend the similar limiting procedure to the double-rotating charged black hole of the five-dimensional
minimal gauged supergravity. Finally, we present our conclusions in Section 5. Throughout this paper,
the thermodynamic quantities without and with a hat represent to belong to the super-entropic and
usual black holes, respectively.

\section{Kerr-Newman-AdS$_4$ super-entropic black hole}

\subsection{Thermodynamic quantities}\label{4dSEKNAdS}

We now present a brief review of thermodynamical properties of a four-dimensional Kerr-Newman-AdS
super-entropic black hole, whose metric and Abelian gauge potential are \cite{PRL115-031101,JHEP0615096}:
\bea
ds^2 &=& -\, \frac{\Delta(r)}{\Sigma}\big(dt -l\sin^2\theta d\phi\big)^2
 +\Sigma\Big[\frac{dr^2}{\Delta(r)} +\frac{d\theta^2}{\sin^2\theta}\Big] \nn \\
&& +\frac{\sin^4\theta}{\Sigma}\big[l\, dt -(r^2 +l^2)d\phi\big]^2 \, , \label{SEKNAdS} \\
A &=& \frac{qr}{\Sigma}(dt -l\sin^2\theta d\phi) \, ,
\eea
where $l$ is the cosmological scale, and
$$ \Delta(r) = (r^2 +l^2)^2/l^2 -2mr +q^2 \, , \quad~~ \Sigma = r^2 +l^2\cos^2\theta \, . $$
Here, $m$ and $q$ are the mass and charge parameters, respectively. The azimuthal coordinate
$\phi$ is noncompact and must be compactified by requiring $\phi \sim \phi +\mu$  with $\mu$
being a dimensionless parameter related to a new chemical potential $K$.

The Bekenstein-Hawking entropy is one quarter of the area of the event horizon
\be
S = \mathcal{A}/4 = \mu(r_+^2 +l^2)/2 \, , \label{S}
\ee
where $r_+$ is the location of the event horizon. The Hawking temperature is proportional to
the surface gravity $\kappa$ on the event horizon
\be
T = \frac{\kappa}{2\pi} = \frac{\p_{r_+}\Delta(r_+)}{4\pi(r_+^2 +l^2)}
 =\frac{2r_+(r_+^2 +l^2) -ml^2}{2\pi l^2(r_+^2 +l^2)} \, . \label{T}
\ee

Note that the super-entropic black hole is rotating with the speed of light at infinity, and on
the event horizon its angular velocity is given by
\be
\Omega = -\, \frac{g_{t\phi}}{g_{\phi\phi}}\Big|_{r = r_+} = \frac{l}{r_+^2 +l^2} \, .
\label{Omega}
\ee

The electric charge $Q$ of the black hole can be computed by using the Gauss' law integral
\be
Q = \frac{1}{4\pi}\int\star F = \frac{\mu q}{2\pi} \, , \label{Q}
\ee
and its corresponding electrostatic potential at the event horizon reads
\be
\Phi = (A_{\mu}\chi^{\mu})|_{r=r_+} = \frac{qr_+}{r_+^2 +l^2}\, , \label{Phi}
\ee
where $\chi = \p_t +\Omega\p_{\phi}$ is the Killing vector normal to the event horizon.

To compute the mass $M$ and the angular momentum $J$, the conformal completion method \cite{PRD73-104036}
or the Abbott-Deser method \cite{NPB195-76} is usually used. Here we adopt the former. The idea
is to perform a conformal transformation on the metric (\ref{SEKNAdS}) to remove the divergence
in the integrals at the boundary (conformal infinity). After taking the $r\to \infty$
limit in the line element $ds^2/r^2$, one obtains the boundary metric
\be
ds_\infty^2 = -\, \Big(\frac{dt}{l} -\sin^2\theta d\phi\Big)^2 +\frac{d\theta^2}{\sin^2\theta}
 +\sin^4\theta d\phi^2 \, .
\ee
Then the conserved charges $\mathcal{Q}[\xi]$ associated with the Killing vector $\xi$ can be
computed by
\be
\mathcal{Q}[\xi] = \frac{l^3}{8\pi}\int_{S_\infty} r\, N^{\alpha} N^{\beta}
 C^{\mu}_{~\alpha\nu\beta}\xi^{\nu}dS_{\mu} \, ,
\ee
where $N^\mu = \big[0, -r^2/l^2, 0, 0\big]$ is the vector normal to the boundary,
$C^{\mu}_{~\alpha\nu\beta}$ is the conformal Weyl curvature tensor, and
\be
dS_t = \sin\theta d\theta d\phi/l
\ee
is the temporal component of the area vector in the three-dimensional conformal boundary. So one
can evaluate the mass $M$ and the angular momentum $J$ as
\be
M = \mathcal{Q}[\p_t] = \frac{\mu m}{2\pi} \, , \quad~~
J = \mathcal{Q}[\p_{\phi}] = \frac{\mu ml}{2\pi} \equiv Ml \, . \label{J}
\ee
It should be mentioned that the angular momentum $J$ can also be directly computed by the Komar integral
, while the mass can be obtained by the Komar integral after the substraction of a divergence
arising from the zero-mass background \cite{PRD89-084007}.

In Refs. \cite{PRL115-031101,JHEP0615096}, it has been shown that the above thermodynamic quantities
obey the extended differential and integral mass formulas simultaneously
\bea
dM &=& TdS +\Omega dJ +VdP +\Phi dQ +Kd\mu \, , \label{FL} \\
M &=& 2(TS +\Omega J -VP) +\Phi Q \, , \label{Smarr}
\eea
with the thermodynamic volume and a new chemical potential
\bea
V &=& \frac{2}{3}\mu r_+(r_+^2 +l^2) \label{V} \\
K &=& \frac{(l^2 -r_+^2)[(r_+^2 +l^2)^2 +q^2l^2]}{8\pi l^2r_+(r_+^2 +l^2)} \, , \label{K}
\eea
being conjugate to the pressure $P = 3/(8\pi l^2)$ and $\mu$, respectively.

\subsection{A new squared-mass formula}\label{SQM}

The Christodoulou-Ruffini squared-mass formula \cite{PRL25-1596,PRD4-3552} was initially found
for the Kerr-Newman black hole. Later it was generalized to the Kerr-Newman-AdS$_4$ black hole
case \cite{CQG17-399}. Now we try to derive a similar squared-mass formula for the
Kerr-Newman-AdS$_4$ super-entropic black hole.

Note that the event horizon equation $\Delta(r_+) = 0$ can be re-expressed as
\be
\frac{S^2}{\pi^2l^2} +Q^2 = \frac{\mu Mr_+}{\pi} \, . \label{SId}
\ee
From Eq. (\ref{S}), one can get $r_+^2 = (2S/\mu) -l^2$. Substituting it into the squared Eq. (\ref{SId})
and using $l^2 = 3/(8\pi P)$  and the chirality condition: $J = Ml$,  then after a little algebra we arrive at an identity
\be
M^2 = \frac{1}{2\mu S}\Big(\frac{8P}{3}S^2 +\pi Q^2\Big)^2 +\frac{\mu J^2}{2S} \, , \label{sqm}
\ee
which is our new Christodoulou-Ruffini-like squared-mass formula for the four-dimensional Kerr-Newman-AdS
super-entropic black hole. Obviously, the parameters $S, J, Q, P$ and $\mu$ of the black hole form a whole
set of energetic extensive parameters for the thermodynamical fundamental functional relation $M = M(S, J,
Q, P, \mu)$. If the chirality condition is not being taken into account, however, see the subsection
below for a discussion about its impact on the actual thermodynamics. It is interesting to note that using this squared-mass formula, one can study conveniently the
black hole chemistry and thermodynamic phase transition of the Kerr-Newman-AdS$_4$ super-entropic black hole.

\subsection{The first law of thermodynamics and the Bekenstein-Smarr mass formula}

Differentiating the above squared-mass formula (\ref{sqm}) yields the conjugate quantities of $S,
J, Q, P$ and $\mu$,   all of which, as was done in Refs. \cite{CQG17-399,PLB608-251,PRD21-884}, are viewed formally as independent
thermodynamical variables at this moment.\footnote{However, in fact they are not completely independent of each
other by virtue of the existence of the chirality condition $J = Ml$. We thank the anonymous referee for pointing this
out to us. A careful discussion about its impact on the mass formulas is presented in the next
subsection.} In doing so, we can arrive at the first law (\ref{FL}) and the Bekenstein-Smarr relation (\ref{Smarr}),
with the conjugate thermodynamic potentials given by the ordinary Maxwell relations as follows. The conjugate quantity of the entropy $S$ is the Hawking temperature
\bea
T &=& \frac{\p M}{\p S}\bigg|_{(J,Q,P,\mu)} = -\, \frac{M}{2S}
 +\frac{8P}{3\mu M}\Big(\frac{8P}{3}S^2 +\pi Q^2\Big) \nn \\
&=& \frac{2r_+(r_+^2 +l^2) -ml^2}{2\pi l^2(r_+^2 +l^2)} \, .
\eea
The angular velocity and the electrostatic potential, which are conjugate to $J$ and $Q$, respectively,
are given by
\bea
\Omega &=& \frac{\p M}{\p J}\bigg|_{(S,Q,P,\mu)} = \frac{\mu J}{2SM} = \frac{l}{r_+^2 +l^2} \, ,\\
\Phi &=& \frac{\p M}{\p Q}\bigg|_{(S,J,P,\mu)}
 = \frac{\pi Q}{\mu SM}\Big(\frac{8P}{3}S^2 +\pi Q^2\Big)
 = \frac{qr_+}{r_+^2 +l^2} \, . \quad
\eea
These three conjugate quantities are entirely identical to those given in Eqs. (\ref{T}), (\ref{Omega})
and (\ref{Phi}). Differentiating the squared-mass formula (\ref{sqm}) with respect to the pressure $P$
and the dimensionless parameter $\mu$, one can get the thermodynamical volume
\bea
V &=& \frac{\p M}{\p P}\bigg|_{(S,J,Q,\mu)} = \frac{4S}{3\mu M}\Big(\frac{8P}{3}S^2 +\pi Q^2 \Big) \nn \\
&=& \frac{2}{3}\mu r_+(r_+^2 +l^2) \, ,
\eea
and a new conjugate variable
\be
K = \frac{\p M}{\p \mu}\bigg|_{(S,J,Q,P)} = \frac{-M^2S +\mu J^2}{2\mu SM}
= \frac{m(l^2 -r_+^2)}{4\pi(r_+^2 +l^2)} \, ,
\ee
These two quantities are the same  as those  obtained in Refs. \cite{PRL115-031101,JHEP0615096}. With the
conjugate variables derived from the squared-mass formula (\ref{sqm}), the first law of thermodynamics
is trivially satisfied whilst the integral Bekenstein-Smarr mass formula is easily checked to be
completely obeyed too. Thus, we have verified that the first law of thermodynamics, the
Bekenstein-Smarr mass formula and the Christodoulou-Ruffini-like squared-mass formula are all
valid for the Kerr-Newman-AdS$_4$ super-entropic black hole.

\subsection{{Remark on the impact of the chirality condition}}

In the last subsection, we have not taken into account the impact of the chirality condition ($J = Ml$) on the
thermodynamical relations of the super-entropic black hole. Now let us make a careful discussion about this
issue. It should be reminded that the super-entropic black hole is obtained by taking the ultra-spinning
limit ($a\to l$), and it is actually degenerate due to three thermodynamical quantities ($M, J, P$) satisfying
the following constraint
\be
P = \frac{3M^2}{8\pi\,J^2} \, , \label{constt}
\ee
which means that they are not independent. Consequently, the first law (\ref{FL}) and the Bekenstein-Smarr
relation (\ref{Smarr}) depict physically a degenerate thermodynamical system. One can adopt any two
of these three variables to describe the genuine thermodynamic properties. However, it is much concise and
simpler to present the thermodynamic relations in terms of the enthalpy $M$ and the pressure $P$. After eliminating
$J$ from the mass formulas in favor of $l^2 = 3/(8\pi\,P)$, the first law (\ref{FL}) and the Bekenstein-Smarr
relation (\ref{Smarr}) degenerate to the following nonstandard forms (their thermodynamic quantities cannot
constitute the ordinary conjugate pairs due to the presence of a factor $(1 -\Omega\, l)$ in front of $dM$
and $M$):
\bea
(1 -\Omega\, l)dM &=& TdS +V^{\prime} dP +\Phi\, dQ +Kd\mu \, , \\
(1 -\Omega\, l)M &=& 2(TS -V^{\prime}P) +\Phi\, Q \, ,
\eea
where
\be
V^{\prime} = V -\frac{J\Omega}{2P} = V -\frac{4\pi}{3}\Omega\, M\, l^3 \, .
\ee
Noting that the lowest two eigenvalues of Virasoro algebra: $L_+ = M$ and $L_- = 0$ when $J = Ml$, the above
expressions reproduce Eqs. (19-21) in Ref. \cite{PRL115-031101} and Eq. (23) in Ref. \cite{PRD89-084007}.

In the same way, by considering that $J$ is a redundant variable (although it is a real measurable quantity),
our squared-mass formula (\ref{sqm}) reduces to
\be
2\mu\,M^2\Big(S -\frac{3\mu}{16\pi\,P}\Big) = \Big(\frac{8P}{3}S^2 +\pi\,Q^2\Big)^2 \, .
\ee
Taking the positive square root of this formula, one gets a fundamental  thermodynamical relation
\be
M\sqrt{2\mu\,S -\frac{3\mu^2}{8\pi\,P}} = \frac{8P}{3}S^2 +\pi\,Q^2 \, ,
\label{mass2}
\ee
which coincides with Eq. (24) given originally in Ref. \cite{PRD89-084007}.

Eq. (\ref{mass2}) suggests that the enthalpy $M$ be viewed as the functional relation $M = M(S, Q, P, \mu)$.
Similar to the strategy as that in the last subsection, the above nonstandard differential and integral mass
formulas can be deduced by exploiting the standard Maxwell rule. Likewise, one perhaps prefers to eliminate
$P$ via Eq. (\ref{constt}) from the beginning, but the resulted expressions would be very complicated, and
will not be presented here. It should be pointed out that the discussion made in this subsection can be
easily generalized to higher dimensions too.

It is interesting to make a comparison of the above discussions with a recent work \cite{PRD100-101501} on the
thermodynamics of four-dimensional Taub-NUT spacetimes, where a new secondary hair $J_n = mn$ is introduced
to perfectly cast their thermodynamics into the standard forms of usual black hole thermodynamics. Without
introducing $J_n = mn$ into the mass formulas, thermodynamical relations would be inconsistent (nonstandard).

\subsection{A simple limiting procedure to obtain the thermodynamic quantities of the Kerr-Newman-AdS$_4$
super-entropic black hole}\label{asm}

In the above subsection, we have presented all thermodynamic quantities of a Kerr-Newman-AdS$_4$
super-entropic black hole. It was claimed in Ref. \cite{PRL115-031101,JHEP0615096,1911.12817} that
these thermodynamic quantities can not be obtained by taking the $a\to l$ limit of the usual
Kerr-Newman-AdS thermodynamic quantities, due to the singular nature of the ultra-spinning limit.
However, we will show that this is not the case and suggest a method that the $a\to l$ limit can
be properly performed. Below, we try to provide a simple way to derive them from those of the usual
Kerr-Newman-AdS$_4$ black hole by taking the ultra-spinning limit appropriately.

It should be reminded that the Kerr-Newman-AdS$_4$ super-entropic black hole is constructed by
taking the $a\to l$ limit of its corresponding usual Kerr-Newman-AdS$_4$ black hole in a frame
rotating at infinity. However, as is pointed out in~\cite{CQG22-1503} that all thermodynamic quantities
of the usual Kerr-Newman-AdS$_4$ black hole that enter  the first law and the Bekenstein-Smarr
mass formula should be those quantities (especially, the conserved mass, the angular velocity at
the horizon relative to the infinity, and the thermodynamic volume) measured in a frame rest at
infinity. Transforming all thermodynamic quantities into a frame that rotates at infinity,  we have
\be
\beal
&\hat{M} = \frac{m}{\Xi} \, , \quad~~ \hat{J} = \frac{ma}{\Xi^2} \, , \quad~~
\hat{Q} = \frac{q}{\Xi} \, , \quad~~ \hat{P} \equiv P = \frac{3}{8\pi l^2} \, , \\
&\hat{T} = \frac{\p_{r_+}\hat{\Delta}(r_+)}{4\pi(r_+^2 +a^2)}
 = \frac{(2r_+^2 +a^2 +l^2)r_+ -ml^2}{2\pi l^2(r_+^2 +a^2)} \, , \\
&\hat{S} = \frac{\pi(r_+^2 +a^2)}{\Xi} \, , \quad~~ \hat{\Omega} = \frac{a\Xi}{r_+^2 +a^2} \, ,
\quad~~ \hat{\Phi} = \frac{qr_+}{r_+^2 +a^2} \, ,
\eeal \label{bhT}
\ee
where
$$ \hat{\Delta}(r) = (1 +r^2/l^2)(r^2 +a^2) -2mr +q^2 \, , \quad~~ \Xi = 1 -a^2/l^2 \, . $$
These thermodynamic quantities still satisfy the Bekenstein-Smarr mass formula
\be
\hat{M} = 2(\hat{T}\hat{S} +\hat{\Omega}\hat{J} -\hat{V}\hat{P}) +\hat{\Phi}\hat{Q} \, ,
\label{BS}
\ee
however, the first law now boils down to a differential identity
\be
d\hat{M} = \hat{T}d\hat{S} +\hat{\Omega}d\hat{J} +\hat{V}d\hat{P} +\hat{\Phi}d\hat{Q}
 +\frac{\hat{J}}{2a}d\Xi \, , \label{MId}
\ee
with the thermodynamic volume
\be
\hat{V} = \frac{4}{3}r_+\hat{S} = \frac{4\pi}{3\Xi}r_+(r_+^2 +a^2) \, .
\ee

We now want to take the $a\to l$ limit of the above thermodynamic quantities with an implicit
assumption that the mass and charge parameters remain unchanged when this limit is taken, and
so the horizon radius reduces to the one after taking the $a\to l$ limit, which is clear from
the horizon equation. Taking straightforwardly the $a\to l$ limit, which means $\Xi\to 0$, then
one can see that the temperature $\hat{T}$ equals to $T$ defined in Eq. (\ref{T}), the electrostatic
potential $\hat{\Phi}$ reduces to $\Phi$ given in Eq. (\ref{Phi}), but $(\hat{M},\hat{J},\hat{Q},
\hat{V})\to \infty$, and $\hat{\Omega}\to 0$. This was first observed in \cite{JHEP0615096} that
the thermodynamic quantities of the super-entropic black hole can not be obtained directly from
those of the usual rotating AdS black hole by taking the $a\to l$ limit.

However, this is only  superficial as can be seen by noting that a coordinate transformation on the
azimuthal coordinate has to be done in the process of obtaining the super-entropic black hole
solution via taking the $a\to l$ limit. That is, before taking the ultra-spinnig limit $a\to l$,
one needs to redefine a new azimuthal coordinate $\phi = \hat{\phi}/\Xi$ ($\hat{\phi}$ has a period
$2\pi$ to prevent a conical singularity) and to identify it with period $\mu$ to avoid a singular
metric in this limit. Only after this coordinate transformation has been done, can the $a\to l$
limit then be taken to obtain a regular super-entropic black hole.

Inspecting into the calculation of the conserved charges by means of the conformal completion method
and also of the entropy via the horizon area, one has to perform the integral about the azimuthal
coordinates ($\hat{\phi}, \phi$ has a respective period $2\pi, \mu$). It is clear that one should
consider the thermodynamic quantities ($\Xi\hat{M}, \hat{\Omega}/\Xi, \Xi^2\hat{J}, \Xi\hat{Q},
\Xi\hat{S}, \Xi\hat{V}$) that are all finite in the $a\to l$ limit. Note that the excess factor
$\Xi^{\pm}$ that appears in the angular momentum and the angular velocity to remove the divergence
or the zero is due to the rescaling of the azimuthal coordinate.

Taking into account of the above consideration, we suggest that the following relations
\be
\beal
&M = \frac{\mu\Xi}{2\pi}\hat{M} \, , \quad~~ J = \frac{\mu\Xi^2}{2\pi}\hat{J} \, , \quad~~
 Q = \frac{\mu\Xi}{2\pi}\hat{Q} \, , \\
&S = \frac{\mu\Xi}{2\pi}\hat{S} \, , \qquad \Omega = \frac{1}{\Xi}\hat{\Omega} \, , \qquad~
V = \frac{\mu\Xi}{2\pi}\hat{V} \, , \\
& T = \hat{T} \, , \qquad\quad~ \Phi = \hat{\Phi} \, , \qquad\quad~ P = \hat{P}
\eeal \label{rels}
\ee
should be established between the thermodynamic quantities of the usual Kerr-Newman-AdS$_4$ black
hole and their corresponding super-entropic ones when taking the $a\to l$ limit.

Substituting the relations (\ref{rels}) into (\ref{BS}) and taking the $a\to l$ limit yields directly
the Bekenstein-Smarr mass formula (\ref{Smarr}). Considering the differential identity (\ref{MId}),
and viewing the dimensional parameter $\mu$ as a thermodynamic variable, we find that the term related to
$d\Xi$ just vanishes properly and the identity perfectly reduces to the first law (\ref{FL}) with
\be
K = \frac{M -\Phi Q -2VP}{2\mu} = \frac{M -TS -\Omega J -\Phi Q}{\mu} \, .
\ee

We now turn to consider the squared-mass formula. To this end, one can deduce the following squared-mass identity:
\be
\frac{\hat{S}}{\pi\Xi}\hat{M}^2 = \hat{J}^2
 +\frac{1}{4\pi^2}\Big(\hat{S} +\frac{8P}{3}\hat{S}^2 +\pi\hat{Q}^2\Big)^2 \, .
\ee
Taking advantage of the relations (\ref{rels}), it becomes
\be
M^2 = \frac{1}{2\mu\,S}\Big(\frac{\mu\Xi}{2\pi}S +\frac{8P}{3}S^2 +\pi\,Q^2\Big)^2
 +\frac{\mu\,J^2}{2S} \, ,
\ee
which reduces to the squared-mass formula (\ref{sqm}) after taking the $a\to l$ limit. Similarly, the
chirality condition ($J = Ml$) can be recovered by the same procedure from the relation: $\hat{J}=\hat{M}a/\Xi$.

The above method is a very simple, effective approach to obtain easily all the thermodynamic quantities
of the super-entropic black hole from their counterparts of the usual Kerr-Newman-AdS$_4$ black hole.

\section{Singly-rotating Kerr-AdS super-entropic black holes in arbitrary dimensions}

To demonstrate that the relations (\ref{rels}) are also applicable to the case of black holes in arbitrary
dimensions, let us now consider the singly-rotating $d$-dimensional Kerr-AdS spacetimes in the
frame rotating at infinity \cite{PRD59-064005}
\be
\beal
ds^2 &= -\, \frac{\bar{\Delta}_r}{\bar\Sigma}\Big(dt -\frac{a}{\Xi}\sin^2\theta d\phi\Big)^2
 +\frac{\bar\Sigma}{\bar{\Delta}_r}dr^2 \\
&\quad +\frac{\bar\Sigma}{\Delta_\theta}d\theta^2
 +\frac{\Delta_\theta\sin^2\theta}{\bar\Sigma}\Big(a\, dt -\frac{r^2 +a^2}{\Xi}d\phi\Big)^2 \\
&\quad +r^2\cos^2\theta\, d\Omega_{d-4}^2 \, ,
\eeal
\ee
where
\bea
&&\bar{\Delta}_r = (r^2 +a^2)(1 +r^2l^{-2}) -2mr^{5-d} \, , \quad~~ \Xi = 1 -a^2l^{-2} \, , \nn \\
&&\Delta_\theta = 1 -a^2l^{-2}\cos^2\theta \, , \quad~~ \bar\Sigma = r^2 +a^2\cos\theta \, . \nn
\eea
The thermodynamic quantities of these black holes have the following expressions in the extended phase
space
\be
\beal
&\hat{M} = \frac{\omega_{d-2}}{8\pi\Xi}(d-2)m \, , \quad~~ \hat{J} = \frac{\omega_{d-2}}{4\pi\Xi^2}ma \, ,
\quad~~ \hat{\Omega} = \frac{a\Xi}{r_+^2 +a^2} \, , \\
&\hat{S} = \frac{\omega_{d-2}}{4\Xi}(r_+^2 +a^2)r_+^{d-4} \, ,
\quad~~ \hat{V} = \frac{\omega_{d-2}}{(d-1)\Xi}(r_+^2 +a^2)r_+^{d-3} \, , \\
&\hat{T} = \frac{(d-1)r_+^4 +(d-3)(a^2 +l^2)r_+^2 +(d-5)a^2l^2}{4\pi r_+(r_+^2 +a^2)l^2} \, ,
\eeal
\ee
where $\omega_{d-2}$ is the volume of the unit ($d-2$)-sphere:
$$ \omega_{d-2} = \frac{2\pi^{\frac{d-1}{2}}}{\Gamma(\frac{d-1}{2})} \, . $$

The above thermodynamic quantities satisfy an extended Bekenstein-Smarr mass formula
\be
(d-3)\hat{M} = (d-2)(\hat{T}\hat{S} +\hat{\Omega}\hat{J} -\hat{V}\hat{P}) \, ,
\ee
however the first law boils down to a differential identity as before
\be
d\hat{M} = \hat{T}d\hat{S} +\hat{\Omega}d\hat{J} +\hat{V}d\hat{P} +\frac{\hat{J}}{2a}d\Xi \, ,
\label{dMId}
\ee
where the pressure conjugate to the thermodynamic volume is
\be
P = -\, \frac{\Lambda}{8\pi} = \frac{(d-1)(d-2)}{16\pi l^2} \, .
\ee

Three steps to construct the super-entropic versions of the higher-dimensional singly-rotating Kerr-AdS
black holes are \cite{PRL115-031101,JHEP0615096}: (1) redefine the angle coordinate $\phi$ by multiplying
it with a factor $\Xi$; (2) take the $a\to l$ limit; (3) compactify the $\phi$-direction with a period
of the dimensional parameter $\mu$. Using the relations (\ref{rels}) and taking the ultra-spinning limit
$a\to l$ (with the assumption that the mass parameter $m$ remains unchanged again), we can easily obtain
the thermodynamic quantities of the singly-rotating Kerr-AdS super-entropic black holes in arbitrary
dimensions \cite{PRL115-031101,JHEP0615096}
\be
\beal
&M = \frac{\mu\omega_{d-2}}{16\pi^2}(d-2)m \, , \quad~~
J = \frac{\mu\omega_{d-2}}{8\pi^2}ml = \frac{2Ml}{d-2} \, , \\
&S = \frac{\mu\omega_{d-2}}{8\pi}(r_+^2 +l^2)r_+^{d-4} \, , \quad~~
T = \frac{(d-1)r_+^2 +(d-5)l^2}{4\pi r_+l^2} \, , \\
&\Omega = \frac{l}{r_+^2 +l^2} \, , \quad~~
V = \frac{\mu\omega_{d-2}}{2\pi(d-1)}(r_+^2 +l^2)r_+^{d-3} \, .
\eeal
\label{TQs}
\ee
In the four-dimensional case ($d=4$) where $\mu = 2\pi$, the above expressions (\ref{TQs}) reduce to those
given in the last section. It is not difficult to check that these thermodynamic quantities satisfy both
the first law of thermodynamics and the Bekenstein-Smarr mass formula
\bea
dM &=& TdS +\Omega dJ +VdP +Kd\mu \, , \\
(d-3)M &=& (d-2)(TS +\Omega J) -2VP \, ,
\eea
where
\be
K = \frac{M -TS -\Omega J}{\mu} = \frac{m\omega_{d-2}(d-2)(l^2-r_+^2)}{8\pi[(d-2)r_+^2 +(3d-10)l^2]} \, .
\ee
is the conjugate quantity of the variable $\mu$. Similar to what has been done in the last section, one
can also derive the above expression for K via taking the $a\to l$ limit in the differential identity (\ref{dMId}).

We now turn to seek a generalization of the Christodoulou-Ruffini-like squared-mass formula (\ref{sqm})
to higher dimensions.

Following the procedure used in the above section, we can rewrite the horizon equation: $(r_+^2 +l^2)^2
-2ml^2r_+^{5-d} = 0$ as
\be
r_+^{d-3} = \frac{16P}{(d-1)\mu\omega_{d-2}^2M}S^2 \, ,
\ee
and then by expressing $r_+$ as a function of $S$, we can finally deduce a new mass formula
\bea
\Big[\frac{16P}{(d-1)\mu\omega_{d-2}M}S^2 \Big]^{\frac{2}{d-3}}
&=& \frac{4S}{\mu\omega_{d-2}}\Big[\frac{16P}{(d-1)\mu\omega_{d-2}M}S^2\Big]^{\frac{4-d}{d-3}} \nn \\
&& -\, \frac{J^2(d-2)^2}{4M^2} \, \label{Mfd}
\eea
for the singly-rotating Kerr-AdS super-entropic black holes in arbitrary dimensions.

Viewing Eq. (\ref{Mfd}) as the fundamental relation for the thermodynamical function $M = M(S, J, P, \mu)$
and differentiating it with respect to its variables as was done in the last section, their conjugate thermodynamic
quantities can be correctly gotten as those given above. This confirms that using the relations (\ref{rels}),
the thermodynamic quantities of the singly-rotating Kerr-AdS super-entropic black holes in arbitrary dimensions
can also be obtained from those of the singly-rotating Kerr-AdS black holes by properly taking the ultra-spinning
limit.

The above discussions can be readily extended to the general case with multiple rotation parameters in higher
dimensions. It has been mentioned \cite{JHEP0615096} that the ultra-spining limit can be taken in one and only
one rotating axis. Here we additionally point out that all of the remaining rotation axes should be put in the
frame rest at infinity in order to ensure that all thermodynamic quantities measured in this frame obey both
the first law and the Bekenstein-Smarr mass formula simultaneously. Otherwise, the ``first law'' will boil
down to a differential identity only. In the next section, we will illustrate this issue by considering an exact
double-rotating black hole solution to the five-dimensional Einstein-Maxwell-Chern-Simons gauged supergravity
theory.

\section{Super-entropic black hole of five-dimensional minimal gauged supergravity}

In this section, we will generalize the relations (\ref{rels}) to the case of a double-rotating charged black hole
in the five-dimensional minimal gauged supergravity and then check the validity of these relations. The solution
to the five-dimensional Einstein-Maxwell-Chern-Simons gauged supergravity is given by \cite{PRL95-161301}
\bea
ds^2 &=& -\frac{\tilde{\Delta}_r}{r^2(r^2 +y^2)}X^2 +(r^2 +y^2)\Big(\frac{r^2\, dr^2}{\tilde{\Delta}_r}
 +\frac{dy^2}{H}\Big) \nn \\
&& +\frac{H}{r^2 +y^2}Y^2 +\frac{1}{r^2y^2}\Big(ab\, Z +\frac{qy^2}{r^2 +y^2}X\Big)^2 \, , \\
A &=& \frac{\sqrt{3}q}{2(r^2 +y^2)}X \, ,
\eea
where
\bea
X &=& \frac{1 -y^2l^{-2}}{\Xi_a\Xi_b}dt -\frac{a(a^2 -y^2)}{(a^2 -b^2)\Xi_a}d\phi
-\frac{b(b^2 -y^2)}{(b^2 -a^2)\Xi_b}d\psi \, , \nn \\
Y &=& \frac{1 +r^2l^{-2}}{\Xi_a\Xi_b}dt -\frac{a(r^2 +a^2)}{(a^2 -b^2)\Xi_a}d\phi
-\frac{b(r^2 +b^2)}{(b^2 -a^2)\Xi_b}d\psi \, , \nn \\
Z &=& \frac{(1 +r^2l^{-2})(1 -y^2l^{-2})}{\Xi_a\Xi_b}dt -\frac{(r^2 +a^2)(a^2 -y^2)}{(a^2 -b^2)a\Xi_a}d\phi \nn \\
&& -\frac{(r^2 +b^2)(b^2 -y^2)}{(b^2 -a^2)b\Xi_b}d\psi \, , \nn \\
\tilde{\Delta}_r &=& (1 +r^2l^{-2})(r^2 +a^2)(r^2 +b^2) -2mr^2 +q^2 +2qab \, , \nn \\
H &=& -\, (y^{-2} -l^{-2})(a^2 -y^2)(b^2 -y^2) \, , \nn \\
\Xi_a &=& 1 -a^2l^{-2} \, , \quad~~ \Xi_b = 1 -b^2l^{-2} \, . \nn
\eea
The above solution is presented in the frame where both the $\phi$-axis and the $\psi$-axis are rest at
infinity \cite{PRL95-161301}. Thermodynamical properties of this solution are consistent only in this
rest frame \cite{PRL95-161301}. When considering the extended phase space of a variable cosmological
constant, their consistent thermodynamic quantities were given in Ref. \cite{PRD80-084009} and
subsequently in Ref. \cite{PRD84-024037}.

We would like to boost the rotating angular velocity of the $\phi$-axis to the speed of light, then
we need in advance to make a coordinate transformation: $\phi\to \phi +a\,t/l^2$ to a special frame
where the $\phi$-axis is rotating whilst the $\psi$-axis is non-rotating at infinity. Transforming the
thermodynamic quantities given in Refs. \cite{PRL95-161301,PRD84-024037,PRD80-084009} into this special
frame, the expressions for the mass, electric charge, two angular momenta, Bekenstein-Hawking entropy,
Hawking temperature, electrostatic potential, and two angular velocities are given as follows:
\be
\beal
&\hat{M} = \frac{\pi(2 +\Xi_b)(m +qabl^{-2})}{4\Xi_a\Xi_b^2} \, , \quad~~
 \hat{Q} = \frac{\sqrt{3}\pi q}{4\Xi_a\Xi_b} \, , \\
&\hat{J}_\phi = \frac{\pi[2ma +qb(2 -\Xi_a)]}{4\Xi_a^2\Xi_b} \, , \\
&\hat{J}_\psi = \frac{\pi[2mb +qa(2 -\Xi_b)]}{4\Xi_a\Xi_b^2} \, , \\
& \hat{S} = \frac{\pi^2[(r_+^2 +a^2)(r_+^2 +b^2) +qab]}{2\Xi_a\Xi_br_+} \, , \\
& \hat{T} = \frac{(2r_+^2 +a^2 +b^2 +l^2)r_+^4 -(q^2 +2qab +a^2b^2)l^2 }{2\pi l^2r_+(r_+^2 +a^2)} \, , \\
& \hat{\Phi} = \frac{\sqrt{3}qr_+^2}{(r_+^2 +a^2)(r_+^2 +b^2) +qab} \, , \\
& \hat{\Omega}_\phi = \frac{\Xi_a[a(r_+^2 +b^2) +qb]}{(r_+^2 +a^2)(r_+^2 +b^2) +qab} \, , \\
& \hat{\Omega}_\psi = \frac{b(1 +r_+^2l^{-2})(r_+^2 +a^2) +qa}{(r_+^2 +a^2)(r_+^2 +b^2) +qab} \, ,
\eeal
\label{TQemcs}
\ee
where $\hat{\Omega}_\phi$ represents the horizon angular velocity around the $\phi$-axis, whilst
$\hat{\Omega}_\psi$ is the one around the $\psi$-axis, which is evaluated at horizon relative to
the infinity.

In this special frame, the above thermodynamic quantities~(\ref{TQemcs}) satisfy the integral
Bekenstein-Smarr mass formula:
\be
2\hat{M} = 3(\hat{T}\hat{S} +\hat{\Omega}_\phi d\hat{J}_\phi
 +\hat{\Omega}_\psi d\hat{J}_\psi) +2(\hat{\Phi}\hat{Q} -\hat{V}\hat{P}) \, ,
\label{IM5}
\ee
whilst the differential mass formula merely represents a differential identity only
\be
d\hat{M} = \hat{T}d\hat{S} +\hat{\Omega}_\phi d\hat{J}_\phi +\hat{\Omega}_\psi d\hat{J}_\psi
 +\hat{\Phi}d\hat{Q} +\hat{V}d\hat{P} +\frac{\hat{J}_\phi}{2a}d\Xi_a \, ,
\label{dM5}
\ee
where the thermodynamic volume $\hat{V}$ conjugate to the pressure $\hat{P} = P = 3/(4\pi l^2)$ has
the form
\bea
\hat{V} &=& \frac{\pi^2}{6\Xi_a\Xi_b}\Big\{3(r_+^2 +a^2)(r_+^2 +b^2) +2qab \nn \\
&& +\frac{b[2mb +qa(2 -\Xi_b)]}{\Xi_b}\Big\} \ .
\eea

Now assuming that the ultra-spinning direction is along the $\phi$-axis of the above special frame, and then
taking the $a\to l$ limit (after defining a new angle coordinate $\phi$ by multiplying the old one
with a factor $\Xi_a$) yields the expected super-entropic black hole solution presented in Ref.
\cite{JHEP0615096}. Then we  take the same ultra-spinning limit $a\to l$ on the above
thermodynamic quantities of the double-rotating charged black hole in the five-dimensional minimal
gauged supergravity theory. Now the relations (\ref{rels}) in the singly-rotating case should be
generalized as follows:
\be
\beal
&M = \frac{\mu\Xi_a}{2\pi}\hat{M} \, , \quad Q = \frac{\mu\Xi_a}{2\pi}\hat{Q} \, , \quad
 J_\phi = \frac{\mu\Xi_a^2}{2\pi}\hat{J}_\phi \, , \\
&J_\psi = \frac{\mu\Xi_a}{2\pi}\hat{J}_\psi \, , \quad
 \Omega_\phi = \frac{1}{\Xi_a}\hat{\Omega}_\phi \, ,
 \quad S = \frac{\mu\Xi_a}{2\pi}\hat{S} \, , \\
&V = \frac{\mu\Xi_a}{2\pi}\hat{V} \, , \quad
 T = \hat{T} \, , \quad \Omega_\psi = \hat{\Omega}_\psi \, , \quad \Phi = \hat{\Phi}
\eeal
\label{grel}
\ee
together with $P = \hat{P}$. Similarly, we have assumed that the mass parameter $m$, the electric
charge parameter $q$, and one rotation parameter $b$ remain unchanged.

Then taking the $a\to l$ limit, one can get straightforwardly the thermodynamic quantities of the
corresponding super-entropic black hole
\be\beal
&M = \frac{\mu(2 +\Xi_b)(m +qb/l)}{8\Xi_b^2} \, , \quad~~ Q = \frac{\sqrt{3}\mu q}{8\Xi_b} \, , \\
&J_\phi = \frac{\mu(ml +qb)}{4\Xi_b} \, , \quad~~ J_\psi = \frac{\mu[2mb +ql(2 -\Xi_b)]}{8\Xi_b^2} \, , \\
&\Omega_\phi = \frac{[l(r_+^2 +b^2) +qb]}{(r_+^2 +l^2)(r_+^2 +b^2) +qbl} \, , \\
&\Omega_\psi = \frac{bl^{-2}(r_+^2 +l^2)^2 +ql}{(r_+^2 +l^2)(r_+^2 +b^2) +qbl} \, , \\
&S = \frac{\pi\mu[(r_+^2 +l^2)(r_+^2 +b^2) +qbl]}{4\Xi_br_+} \, , \\
&T = \frac{r_+^4[2 +(2r_+^2 +b^2)l^{-2}] -(q +bl)^2}{2\pi r_+[(r_+^2 +l^2)(r_+^2 +b^2) +qbl]} \, , \\
&\Phi = \frac{\sqrt{3}qr_+^2}{(r_+^2 +l^2)(r_+^2 +b^2) +qbl} \, , \\
&V = \frac{\pi\mu}{12\Xi_b}\Big\{3(r_+^2 +l^2)(r_+^2 +b^2) +2qbl \nn \\
&\qquad +\frac{b[2mb +ql(2 -\Xi_b)]}{\Xi_b}\Big\} \, .
\eeal\label{bhT2}
\ee
These expressions are the same ones as those initially obtained in Ref. \cite{JHEP0615096}. They satisfy
both the differential and integral mass formulas simultaneously
\bea
dM &=& TdS +\Omega_\phi dJ_\phi +\Omega_\psi dJ_\psi +\Phi dQ +VdP \, , \\
2M &=& 3(TS +\Omega_\phi dJ_\phi +\Omega_\psi dJ_\psi) +2(\Phi Q -VP) \, ,
\eea
if $\mu$ is really a constant. Dealing with the integral mass formula (\ref{IM5}) and the differential
identity (\ref{dM5}) for a constant $\mu$ also leads to the above first law and the Bekenstein-Smarr
mass formula. Therefore, our limiting procedure is also applicable to coping with the case of black
holes carrying multiple rotation parameters as well.

\section{Conclusions}

The super-entropic black hole has spurred an increasing deal of recent interest, due to the discovery that it
has a noncompact horizon topology and violates the ``reverse isoperimetric inequality", but its thermodynamic
quantites satisfy both the first law and the Bekenstein-Smarr mass formula. In this paper, we obtain firstly
a new Christodoulou-Ruffini-like squared-mass formula for the four-dimensional Kerr-Newman-AdS super-entropic
black hole and its extension to higher dimensions with just one rotation parameter. Differentiating the
squared-mass formulas yields the conjugate partners of their corresponding thermodynamic variables, which
satisfy both the first law and the Bekenstein-Smarr mass formula.  Then the impact of the chirality condition on the actual thermodynamics is discussed. After that, 
 we construct a set of very simple
relations between thermodynamic quantities of the usual Kerr-Newman-AdS$_4$ black hole and those of its
super-entropic counterpart. Using these relations, we find that the thermodynamic quantities of the
super-entropic Kerr-Newman-AdS$_4$ black hole can be derived from those of its corresponding usual black
hole by taking the ultra-spinning limit appropriately. Our method is then generalized to the singly-rotating
Kerr-AdS black hole in arbitrary dimensions and the double-rotating charged black hole of the five-dimensional
minimal gauged supergravity. From our discussions completed in this article, it is natural to infer that our
method can be used to obtain the thermodynamic quantities of all super-entropic black holes from those of
their usual counterparts by taking the ultra-spinning limit properly and is in accordance with the spirit
of obtaining these solutions by taking the same limit.

\acknowledgments

This work is supported by the National Natural Science Foundation of China (NSFC) under Grant
No. 11775077, No. 11690034 ,No. 11435006, No. 11675130 and No. 11275157, and by the Science
and Technology Innovation Plan of Hunan province under Grant No. 2017XK2019.


\begin{thebibliography}{99}

\bibitem{PRL115-031101}
R.A. Hennigar, R.B. Mann, and D. Kubiz\v{n}\'ak,
{Entropy inequality violations from ultraspinning black holes},
\href{http://dx.doi.org/10.1103/PhysRevLett.115.031101}
{Phys. Rev. Lett. \textbf{115}, 031101 (2015)}.

\bibitem{PRD89-084007}
D. Klemm,
{Four-dimensional black holes with unusual horizons},
\href{http://dx.doi.org/10.1103/PhysRevD.89.084007}
{Phys. Rev. D \textbf{89}, 084007 (2014)}.

\bibitem{PRD84-024037}
M. Cveti\v{c}, G.W. Gibbons, D. Kubiz\v{n}\'ak, and C.N. Pope,
{Black hole enthalpy and an entropy inequality for the thermodynamic volume},
\href{http://dx.doi.org/10.1103/PhysRevD.84.024037}
{Phys. Rev. D \textbf{84}, 024037 (2011)}.

\bibitem{PRD87-104017}
B.P. Dolan, D. Kastor, D. Kubiz\v{n}\'ak, R.B. Mann, and J. Traschen,
{Thermodynamic volumes and isoperimetric inequalities for de Sitter black holes},
\href{http://dx.doi.org/10.1103/PhysRevD.87.104017}
{Phys. Rev. D \textbf{87}, 104017 (2013)}.

\bibitem{JHEP0615096}
R.A. Hennigar, D. Kubiz\v{n}\'ak, R.B. Mann, and N. Musoke,
{Ultraspinning limits and super-entropic black holes},
\href{http://dx.doi.org/10.1007/JHEP06(2015)096}
{J. High Energy Phys. \textbf{1506}, 096 (2015)}.

\bibitem{NPB903-400}
R.A. Hennigar, D. Kubiz\v{n}\'ak, R.B. Mann, and N. Musoke,
{Ultraspinning limits and rotating hyperboloid membranes},
\href{http://dx.doi.org/10.1016/j.nuclphysb.2015.12.017}
{Nucl. Phys. B \textbf{903}, 400 (2016)}.

\bibitem{PRD95-046002}
S.M. Noorbakhsh and M. Ghominejad,
{Ultra-spinning gauged supergravity black holes and their Kerr/CFT correspondence},
\href{http://dx.doi.org/10.1103/PhysRevD.95.046002}
{Phys. Rev. D \textbf{95}, 046002 (2017)}.

\bibitem{JHEP0118042}
S.M. Noorbakhsh and M.H. Vahidinia,	
{Extremal vanishing horizon Kerr-AdS black holes at ultraspinning limit},
\href{http://dx.doi.org/10.1007/JHEP01(2018)042}
{J. High Energy Phys. \textbf{1801}, 042 (2018)}.

\bibitem{1702.03448}
S.M. Noorbakhsh and M. Ghominejad,
{Higher dimensional charged AdS black holes at ultra-spinning limit and their 2d CFT duals},
\href{http://arxiv.org/abs/arXiv:1702.03448}{arXiv:1702.03448}.

\bibitem{1911.12817}
M. Appels, L. Cuspinera, R. Gregory, P. Krtou\v{s}, and D. Kubiz\v{n}\'ak,
{Are Superentropic black holes superentropic?},
\href{http://arxiv.org/abs/arXiv:1911.12817}{arXiv:1911.12817}.

\bibitem{PRD7-2333}
J.D. Bekenstein,
{Black holes and entropy},
\href{http://dx.doi.org/10.1103/PhysRevD.7.2333}
{Phys. Rev. D \textbf{7}, 2333 (1973)}.

\bibitem{PRD13-191}
S.W. Hawking,
{Black holes and thermodynamics},
\href{http://dx.doi.org/10.1103/PhysRevD.13.191}
{Phys. Rev. D \textbf{13}, 191 (1976)}.

\bibitem{PRL30-71}
L. Smarr,
{Mass formula for Kerr black holes},
\href{http://dx.doi.org/10.1103/PhysRevLett.30.71}
{Phys. Rev. Lett. \textbf{30}, 71 (1973)};
\href{http://dx.doi.org/10.1103/PhysRevLett.30.521}
{\textbf{30}, 521(E) (1973)}.

\bibitem{PRL25-1596}
D. Christodoulou,
{Reversible and irreversible transforations in black hole physics},
\href{http://dx.doi.org/10.1103/PhysRevLett.25.1596}
{Phys. Rev. Lett. \textbf{25}, 1596 (1970)}.

\bibitem{PRD4-3552}
D. Christodoulou and R. Ruffini,
{Reversible transformations of a charged black hole},
\href{http://dx.doi.org/10.1103/PhysRevD.4.3552}
{Phys. Rev. D \textbf{4}, 3552 (1971)}.

\bibitem{PRD100-101501}
S.Q. Wu and D. Wu,
{Thermodynamical hairs of the four-dimensional Taub-Newman-Unti-Tamburino spacetimes},
\href{http://dx.doi.org/10.1103/PhysRevD.100.101501}
{Phys. Rev. D \textbf{100}, 101501 (2019)}.

\bibitem{CQG26-195011}
D. Kastor, S. Ray, and J. Traschen,	
{Enthalpy and the mechanics of AdS black holes},
\href{http://dx.doi.org/10.1088/0264-9381/26/19/195011}
{Classical Quantum Gravity \textbf{26}, 195011 (2009)}.

\bibitem{PRD92-124069}
A.M. Frassino, R.B. Mann, and J.R. Mureika,
{Lower-dimensional black hole chemistry},
\href{http://dx.doi.org/10.1103/PhysRevD.92.124069}
{Phys. Rev. D \textbf{92}, 124069 (2015)}.

\bibitem{CQG34-063001}
D. Kubiz\v{n}\'ak, R.B. Mann, and M. Teo,
{Black hole chemistry: thermodynamics with Lambda},
\href{http://dx.doi.org/10.1088/1361-6382/aa5c69}
{Classical Quantum Gravity \textbf{34}, 063001 (2017)}.

\bibitem{RMP67-605}
G. Ruppeiner,
{Riemannian geometry in thermodynamic fluctuation theory},
\href{http://dx.doi.org/10.1103/RevModPhys.67.605}
{Rev. Mod. Phys. \textbf{67}, 605 (1995)};
\href{http://dx.doi.org/10.1103/RevModPhys.68.313}
{\textbf{68}, 313(E) (1996)}.

\bibitem{PRD75-024037}
G. Ruppeiner,
{Stability and fluctuations in black hole thermodynamics},
\href{http://dx.doi.org/10.1103/PhysRevD.75.024037}
{Phys. Rev. D \textbf{75}, 024037 (2007)}.

\bibitem{PRD78-024016}
G. Ruppeiner,
{Thermodynamic curvature and phase transitions in Kerr-Newman black holes},
\href{http://dx.doi.org/10.1103/PhysRevD.78.024016}
{Phys. Rev. D \textbf{78}, 024016 (2008)}.

\bibitem{1909.03887}
S.W. Wei, Y.X. Liu, and R.B. Mann,
{Ruppeiner geometry, phase transitions, and the microstructure of charged AdS black holes},
\href{http://dx.doi.org/10.1103/PhysRevD.100.124033}
{Phys. Rev. D \textbf{100}, 124033 (2019)}.

\bibitem{PRL123-071103}
S.W. Wei, Y.X. Liu, and R.B. Mann,
{Repulsive interactions and universal properties of charged anti-de Sitter black hole microstructures},
\href{http://dx.doi.org/10.1103/PhysRevLett.123.071103}
{Phys. Rev. Lett. \textbf{123}, 071103 (2019)}.

\bibitem{1909.11911}
S.W. Wei and Y.X. Liu,
{Null geodesics, quasinormal modes, and thermodynamic phase transition for
charged black holes in asymptotically flat and dS spacetimes},
\href{https://arxiv.org/abs/arXiv:1909.11911}{arXiv:1909.11911}.

\bibitem{1910.07874}
P. Wang, H.W. Wu, and H.T. Yang,
{Thermodynamic geometry of AdS black holes and black holes in a cavity},
\href{https://arxiv.org/abs/arXiv:1910.07874}{arXiv:1910.07874}.

\bibitem{1910.03378}
Z.M. Xu, B. Wu, and W.L. Yang,
{The fine micro-thermal structures for the Reissner-Nordstr\"{o}m black hole},
\href{https://arxiv.org/abs/arXiv:1910.03378}{arXiv:1910.03378}.

\bibitem{1910.04528}
S.W. Wei and Y.X. Liu,
{Intriguing microstructures of five-dimensional neutral Gauss-Bonnet AdS black hole},
\href{https://arxiv.org/abs/arXiv:1910.04528}{arXiv:1910.04528}.

\bibitem{PRD73-104036}
W. Chen, H. L\"u, and C. N. Pope,
{Mass of rotating black holes in gauged supergravities},
\href{http://dx.doi.org/10.1103/PhysRevD.73.104036}
{Phys. Rev. D \textbf{73}, 104036 (2006)}.

\bibitem{NPB195-76}
L. F. Abbott and S. Deser,
{Stability of gravity with a cosmological constant},
\href{http://dx.doi.org/10.1016/0550-3213(82)90049-9}
{Nucl. Phys. B \textbf{195}, 76 (1982)}.

\bibitem{CQG17-399}
M.M. Caldarelli, G. Cognola, and D. Klemm,
{Thermodynamics of Kerr-Newman-AdS black holes and conformal field theories},
\href{http://dx.doi.org/10.1088/0264-9381/17/2/310}
{Classical Quantum Gravity \textbf{17}, 399 (2000)}.

\bibitem{PLB608-251}
S.Q. Wu,
New formulation of the first law of black hole thermodynamics: A stringy analogy,
\href{http://dx.doi.org/10.1016/j.physletb.2005.01.018}
{Phys. Lett. B \textbf{608}, 251 (2005)}.

\bibitem{PRD21-884}
D.C. Wright,
Black holes and the Gibbs-Duhem relation,
\href{http://dx.doi.org/10.1103/PhysRevD.21.884}
{Phys. Rev. D \textbf{21}, 884 (1980)}.


\bibitem{CQG22-1503}
G.W. Gibbons, M.J. Perry, and C.N. Pope,
{The first law of thermodynamics for Kerr-anti-de Sitter black holes},
\href{http://dx.doi.org/10.1088/0264-9381/22/9/002}
{Classical Quantum Gravity \textbf{22}, 1503 (2005)}.

\bibitem{PRD59-064005}
S.W. Hawking, C.J. Hunter, and M.M. Taylor-Robinson,
{Rotation and the AdS/CFT correspondence},
\href{http://dx.doi.org/10.1103/PhysRevD.59.064005}
{Phys. Rev. D \textbf{59}, 064005 (1999)}.

\bibitem{PRL95-161301}
Z.W. Chong, M. Cveti\v{c}, H. L\"u, and C.N. Pope,
{General non-extremal rotating black holes in minimal five-dimensional gauged supergravity},
\href{http://dx.doi.org/10.1103/PhysRevLett.95.161301}
{Phys. Rev. Lett. \textbf{95}, 161301 (2005)}.

\bibitem{PRD80-084009}
S.Q. Wu,
{Separability of massive field equations for spin-0 and spin-1/2 charged particles in the general non-extremal rotating charged black holes in minimal five-dimensional gauged supergravity},
\href{http://dx.doi.org/10.1103/PhysRevD.80.084009}
{Phys. Rev. D \textbf{80}, 084009 (2009)}.

\end{thebibliography}
\end{document}